\documentclass{jpsj3}
\usepackage{txfonts}
\usepackage{bm}
\usepackage{enumerate}

\newtheorem{definition}{Definition}
\newtheorem{proposition}{Proposition}
\newtheorem{lemma}{Lemma}
\newtheorem{theorem}{Theorem}
\newtheorem{corollary}{Corollary}

\title{Unbeatable Tit-for-Tat as a Zero-Determinant Strategy}

\author{Masahiko Ueda$^1$\thanks{m.ueda@yamaguchi-u.ac.jp}}
\inst{$^1$Graduate School of Sciences and Technology for Innovation, Yamaguchi University, Yamaguchi 753-8511, Japan} 

\abst{Tit-for-Tat strategy is a strategy in repeated two-player symmetric games which imitates the previous action of the opponent.
We show that the Tit-for-Tat strategy is a zero-determinant strategy, which unilaterally equalizes the expected payoffs of two players, if and only if the stage game is a potential game.
Because it has been known that this condition is equivalent to the condition that the Tit-for-Tat strategy is unbeatable, our results suggest some relation between unbeatable property and the concept of zero-determinant strategy.
}

\begin{document}
\maketitle

\section{Introduction}
\label{sec:intro}
The Tit-for-Tat (TFT) strategy was discovered as a cooperative strategy in the infinitely repeated prisoner's dilemma game, which imitates the previous action of the opponent \cite{RCO1965}.
Although a pair of TFT does not form subgame perfect equilibrium, it forms Nash equilibrium.
Axelrod obtained the numerical results that TFT is the most successful strategy in the prisoner's dilemma game by using computer tournaments \cite{AxeHam1981}.
Meanwhile, in evolutionary games, it was pointed out that TFT is not successful because it is not robust against errors \cite{NowSig1992,NowSig1993,IFN2007}.
Recently, it was found that TFT is contained in the class of zero-determinant (ZD) strategies, which unilaterally enforce linear relations between expected payoffs \cite{PreDys2012}.
Furthermore, it was shown that TFT is also a deformed ZD strategy which unilaterally equalizes all moments of payoffs of two players \cite{Ued2021a}.
Variants of TFT were recently proposed which are robust against implementation errors \cite{YBC2017,MurBae2020}.

Although many researchers investigated properties of TFT in the repeated prisoner's dilemma games, little is known about properties of TFT in other repeated two-player symmetric games.
Although simple, imitation strategies are generally successful in several situations \cite{Veg1997,Sch1998,DOS2012b,BaiNai2018}.
Recently, Duersch et al. found that TFT is unbeatable if and only if the stage game is a potential game \cite{DOS2014}.
Potential games are a class of games in strategic form which have potential functions \cite{MonSha1996}.
For potential games, Nash equilibrium is realized as the maximum of a potential.
Potential games contain several important situations such as the Cournot duopoly game and the public goods game, in addition to the prisoner's dilemma game.
For non-potential games, such as rock-paper-scissors game, TFT can be exploited unboundedly.

In this paper, we show that, in infinitely repeated two-player symmetric games, TFT is unbeatable if and only if TFT is a ZD strategy.
ZD strategies attract much attention because of their counterintuitive properties that the payoffs are unilaterally controlled by one player \cite{HRZ2015,HTS2015,HDND2016,McAHau2017,Ued2021b}.
We show that TFT is a ZD strategy, which unilaterally equalizes the expected payoffs of two players, if and only if the stage game is a potential games, even if the stage game is not the prisoner's dilemma game.
When combined with the results of Ref. \cite{DOS2014}, we can see that the unbeatable property of TFT is equivalent to that TFT is a ZD strategy.

This paper is organized as follows.
In Section \ref{sec:model}, we introduce a model of infinitely repeated two-player symmetric games.
In Section \ref{sec:preliminaries}, we introduce basic concepts used in the later sections and results of the previous papers \cite{PreDys2012,MonSha1996}.
In Section \ref{sec:results}, we prove our main theorem that TFT is a ZD strategy if and only if the stage game is a potential game.
In addition, we also show that TFT unilaterally equalizes the expected payoffs of two players in potential games.
Moreover, we show that TFT cannot unilaterally enforce any linear relations between expected payoffs in non-potential games, if the opponent uses memory-one strategies.
In Section \ref{sec:example}, we check the main result in two examples.
In Section \ref{sec:discussion}, we introduce the results of Ref. \cite{DOS2014}, and discuss the relation between our results and the results of Ref. \cite{DOS2014}.
In this section, we also provide the results about other imitation strategies.
Section \ref{sec:conclusion} is devoted to concluding remarks.

\section{Model}
\label{sec:model}
We consider a two-player symmetric game.
The set of player is $N:=\{ 1, 2 \}$.
The set of action of player $a$ in the stage game is $A_a = A := \{ 1, \cdots, M \}$, where $M$ is a natural number representing the number of action.
The action of player $a$ is written as $\sigma_a \in A$.
We collectively write $\bm{\sigma}:=\left( \sigma_1, \sigma_2 \right)$, and call $\bm{\sigma}$ a state.
The payoff of player $a\in \{ 1, 2 \}$ in the stage game when the state is $\bm{\sigma}$ is described as $s_a\left( \bm{\sigma} \right)$.
Therefore, the stage game is described as $G:=\left( N, \{ A_a \}_{a\in N}, \{ s_a \}_{a\in N} \right)$ \cite{FudTir1991}.
We introduce the notation that $-a := N \backslash \{a\}$.
We assume that the game is symmetric, that is, $s_2(\sigma_1, \sigma_2)=s_1(\sigma_2, \sigma_1) \quad (\forall \sigma_1, \forall \sigma_2)$.

We repeat the stage game $G$ infinitely.
We write an action of player $a$ at round $t\geq 1$ as $\sigma_a^{(t)}$.
We also introduce the notation $h_{\left[ t:t^\prime \right]} := \left( \bm{\sigma}^{(t)}, \cdots, \bm{\sigma}^{(t^\prime)} \right)$ for $t\leq t^\prime$, and call $h_{\left[ t:t^\prime \right]}$ the history in time interval $\left[ t:t^\prime \right]$.
A strategy of player $a$ in the infinitely repeated game is defined by $\left\{ T_a^{(t)} \left( \sigma_a^{(t)} | h_{[1:t-1]} \right) \right\}_{t=1}^\infty$, where $T_a^{(t)} \left( \sigma_a^{(t)} | h_{[1:t-1]} \right)$ is the conditional probability of taking action $\sigma_a^{(t)}$ at round $t$ when the history is $h_{[1:t-1]}$.
We write the expectation of the quantity $B$ with respect to strategies of both players by $\mathbb{E}[B]$.
The payoff of player $a$ in the infinitely repeated game is defined by
\begin{eqnarray}
 \mathcal{S}_a &:=& (1-\delta) \mathbb{E} \left[ \sum_{t=1}^\infty \delta^{t-1} s_a\left( \bm{\sigma}^{(t)} \right) \right],
\end{eqnarray}
where $\delta$ is a discounting factor satisfying $0\leq \delta \leq 1$.

Below we consider only the case $\delta=1$, where the payoff of player $a$ is described as
\begin{eqnarray}
 \mathcal{S}_a &=& \lim_{T\rightarrow \infty} \frac{1}{T} \mathbb{E} \left[ \sum_{t=1}^T s_a\left( \bm{\sigma}^{(t)} \right) \right].
\end{eqnarray}

\section{Preliminaries}
\label{sec:preliminaries}
In this section, we introduce several concepts used in later sections.
Below, the quantity $\delta_{\sigma, \sigma^\prime}$ represents the Kronecker delta.
We also define $s_0\left( \bm{\sigma} \right):=1$ $(\forall \bm{\sigma})$.

First, we introduce time-independent memory-$n$ strategies.
\begin{definition}
\label{def:TImn}
A strategy of player $a$ is a \emph{time-independent memory-$n$ strategy} $(n\geq 0)$ when it is written in the form
\begin{eqnarray}
 T_a^{(t)}\left( \sigma_a^{(t)} | h_{[1:t-1]} \right) &=& T_a\left( \sigma_a^{(t)} | h_{[t-n:t-1]} \right) \quad \left( \forall \sigma_a^{(t)}, \forall h_{[1:t-1]} \right)
\end{eqnarray}
for all $t\geq n+1$ with some common conditional probability $T_a$.
\end{definition}
It should be noted that, in order to define a strategy, the initial condition for $t\leq n$ must also be given aside from $T_a$.

As a special time-independent memory-one strategy, we introduce the Tit-for-Tat strategy.
\begin{definition}
\label{def:TFT}
A time-independent memory-one strategy of player $a$ is the \emph{Tit-for-Tat (TFT) strategy} when $T_a$ in Definition \ref{def:TImn} is written in the form
\begin{eqnarray}
 T_a\left( \sigma_a | \bm{\sigma}^{\prime} \right) &=& \delta_{\sigma_a, \sigma^\prime_{-a}} \quad \left( \forall \sigma_a, \forall \bm{\sigma}^{\prime} \right).
\end{eqnarray}
\end{definition}
That is, TFT imitates the action of the opponent in the previous round.

Next, we introduce zero-determinant strategies.
For time-independent memory-one strategies $T_a$ of player $a$, we first introduce the Press-Dyson vectors \cite{Aki2016,UedTan2020}
\begin{eqnarray}
 \hat{T}_a\left( \sigma_a | \bm{\sigma}^{\prime} \right) &:=& T_a\left( \sigma_a | \bm{\sigma}^{\prime} \right) -  \delta_{\sigma_a, \sigma^{\prime}_a} \quad \left( \forall \sigma_a, \forall \bm{\sigma}^{\prime} \right).
 \label{eq:PD}
\end{eqnarray}
Because the second term in the right-hand side of Eq. (\ref{eq:PD}) can be regarded as the strategy ``Repeat'', which repeats his/her own action in the previous round, the Press-Dyson vectors are interpreted as the difference between his/her own strategy and ``Repeat''.
By using the Press-Dyson vectors, we define the zero-determinant strategies.
\begin{definition}
\label{def:ZDS}
A time-independent memory-one strategy of player $a$ is a \emph{zero-determinant (ZD) strategy} when its Press-Dyson vectors can be written in the form
\begin{eqnarray}
 \sum_{\sigma_a} c_{\sigma_a} \hat{T}_a\left( \sigma_a | \bm{\sigma}^{\prime} \right) &=& \sum_{b=0}^2 \alpha_{b} s_{b} \left( \bm{\sigma}^{\prime} \right) \quad \left( \forall \bm{\sigma}^{\prime} \right)
 \label{eq:ZDS}
\end{eqnarray}
with some nontrivial coefficients $\left\{ \alpha_{b} \right\}$ and $\left\{ c_{\sigma_a} \right\}$ (that is, not $\alpha_0=\alpha_1=\cdots=\alpha_N=0$, and not $c_1=\cdots=c_M=\mathrm{const.}$).
\end{definition}
In other words, in ZD strategies, a linear combination of the Press-Dyson vectors is described as a linear combination of payoff vectors and a vector of all ones.
We remark that the definition of ZD strategies of player $a$ does not depend on the length of memory of strategies of player $-a$.

In order to see properties of ZD strategies, we first remember that the joint probability of states satisfies the recursion relation
\begin{eqnarray}
 P\left( h_{[1:t+1]} \right) &=& \left\{ \prod_a T_a^{(t+1)}\left( \sigma_a^{(t+1)} | h_{[1:t]} \right) \right\} P\left( h_{[1:t]} \right).
 \label{eq:recursion}
\end{eqnarray}
We also define probability distribution of $\bm{\sigma}^{(t)}$ by
\begin{eqnarray}
 P_t \left( \bm{\sigma}^{(t)} \right) &:=& \sum_{h_{[1:t-1]}} P\left( h_{[1:t]} \right).
\end{eqnarray}
We consider the situation that player $a$ uses a ZD strategy.
By taking $\sum_{\sigma_{-a}^{(t+1)}} \sum_{h_{[1:t]}}$ in both sides of Eq. (\ref{eq:recursion}), we obtain
\begin{eqnarray}
 \sum_{\bm{\sigma}^\prime} \delta_{\sigma^\prime_a, \sigma^{(t+1)}_a} P_{t+1} \left( \bm{\sigma}^\prime \right) &=& \sum_{\bm{\sigma}^\prime} T_a\left( \sigma_a^{(t+1)} | \bm{\sigma}^\prime \right) P_{t} \left( \bm{\sigma}^\prime \right)
\end{eqnarray}
Then, by replacing $\sigma_a^{(t+1)} \rightarrow \sigma_a$ and calculating $\lim_{T\rightarrow \infty} \frac{1}{T} \sum_{t=1}^T$ of both sides, we obtain
\begin{eqnarray}
 \sum_{\bm{\sigma}^\prime} \delta_{\sigma^\prime_a, \sigma_a} P^* \left( \bm{\sigma}^\prime \right) &=& \sum_{\bm{\sigma}^\prime} T_a\left( \sigma_a | \bm{\sigma}^\prime \right) P^* \left( \bm{\sigma}^\prime \right),
\end{eqnarray}
where we have defined the limit distribution
\begin{eqnarray}
 P^* \left( \bm{\sigma} \right) &:=& \lim_{T\rightarrow \infty} \frac{1}{T} \sum_{t=1}^T P_{t} \left( \bm{\sigma} \right).
\end{eqnarray}
This fact is known as Akin's lemma:
\begin{lemma}[\cite{Aki2016,UedTan2020}]
\label{lemma:Akin}
The Press-Dyson vectors satisfy
\begin{eqnarray}
 \sum_{\bm{\sigma}^\prime} P^* \left( \bm{\sigma}^{\prime} \right) \hat{T}_a\left( \sigma_a | \bm{\sigma}^{\prime} \right) &=& 0 \quad (\forall \sigma_a).
\end{eqnarray}
\end{lemma}
We also remark that the payoffs in the repeated games are described by expectation with respect to the limit distribution:
\begin{eqnarray}
 \mathcal{S}_b &=& \lim_{T\rightarrow \infty} \frac{1}{T} \sum_{t=1}^T \sum_{\bm{\sigma}} s_b\left( \bm{\sigma} \right) P_{t} \left( \bm{\sigma} \right) \nonumber \\
 &=& \sum_{\bm{\sigma}} s_b\left( \bm{\sigma} \right) P^* \left( \bm{\sigma} \right).
\end{eqnarray}
Below we write the expectation of quantity $B$ with respect to the limit distribution $P^*$ as $\left\langle B \right\rangle^*$.
Therefore, $\mathcal{S}_b = \left\langle s_b \right\rangle^*$.
The following proposition about ZD strategies is a direct consequence of Akin's lemma.
\begin{proposition}[\cite{PreDys2012,UedTan2020}]
\label{prop:ZDS}
A ZD strategy (\ref{eq:ZDS}) unilaterally enforces a linear relation between expected payoffs:
\begin{eqnarray}
 0 &=& \sum_{b=0}^2 \alpha_{b} \left\langle s_{b} \right\rangle^{*}.
\end{eqnarray}
\end{proposition}
In other words, ZD strategies unilaterally control expected payoffs.
We emphasize that this property of ZD strategies hold regardless of the strategy of player $-a$.
Although we above consider the situation that the action set $A$ is countable, ZD strategies were also extended to games with uncountable action set \cite{McAHau2016}.
For such cases, the argument about probability is replaced by that about probability density, and the Kronecker delta is replaced by the Dirac delta.
Furthermore, we also remark that the concept of ZD strategies was recently extended to memory-$n$ strategies with $n\geq 1$ \cite{Ued2021b,Ued2020}.

Finally, we introduce the concept of potential game \cite{MonSha1996}.
\begin{definition}
\label{def:potential}
A game $G=\left( N, \{ A_a \}_{a\in N}, \{ s_a \}_{a\in N} \right)$ is an \emph{(exact) potential game} when there exist a common function $\Phi(\bm{\sigma})$ satisfying
\begin{eqnarray}
 s_a(\sigma_a, \sigma_{-a}) - s_a(\sigma^\prime_a, \sigma_{-a}) &=& \Phi(\sigma_a, \sigma_{-a}) - \Phi(\sigma^\prime_a, \sigma_{-a}) \quad (\forall \sigma_a, \forall \sigma^\prime_a, \forall \sigma_{-a})
 \label{eq:potential}
\end{eqnarray}
for all player $a$.
\end{definition}
The function $\Phi$ is called a potential function.
Because the Nash equilibrium $\bm{\sigma}^*$ is defined by the condition
\begin{eqnarray}
 s_a(\sigma_a^*, \sigma_{-a}^*) &\geq& s_a(\sigma_a, \sigma_{-a}^*) \quad (\forall a, \forall \sigma_a),
\end{eqnarray}
the condition of the Nash equilibrium for potential games is rewritten as
\begin{eqnarray}
 \Phi (\sigma_a^*, \sigma_{-a}^*) &\geq& \Phi (\sigma_a, \sigma_{-a}^*) \quad (\forall a, \forall \sigma_a).
\end{eqnarray}
Therefore, for a potential game, the Nash equilibrium is realized as the maximum of a potential function.
It should be remarked that the concept of potential game is also defined for the case that the action set $A$ is uncountable.

\section{Results}
\label{sec:results}
In the prisoner's dilemma game, it is known that TFT is a ZD strategy, which unilaterally enforces $\left\langle s_1 \right\rangle^{*} = \left\langle s_2 \right\rangle^{*}$ \cite{PreDys2012}.
A natural question is ``Is TFT also a ZD strategy in other two-player symmetric games?''.
In this section, we show that TFT in two-player symmetric games becomes a ZD strategy if and only if the stage game is a potential game.

We consider the situation that player $1$ takes TFT.
Below, for quantities $B(\sigma_1, \sigma_2)$, we use the following notations
\begin{eqnarray}
 B^{(\mathrm{S})}(\sigma_1, \sigma_2) &:=& \frac{1}{2} \left[ B(\sigma_1, \sigma_2) + B(\sigma_2, \sigma_1) \right] \\
 B^{(\mathrm{A})}(\sigma_1, \sigma_2) &:=& \frac{1}{2} \left[ B(\sigma_1, \sigma_2) - B(\sigma_2, \sigma_1) \right],
\end{eqnarray}
which correspond to symmetric and anti-symmetric parts of $B(\sigma_1, \sigma_2)$, respectively.

\subsection{When is TFT a zero-determinant strategy?}
\label{subsec:potential}
We first prove the following lemma, which is essentially the same as one in Ref. \cite{DOS2012a}.
\begin{lemma}
\label{lemma:potential}
For two-player symmetric games, the definition of potential game is equivalent to the condition
\begin{eqnarray}
 s_1^{(\mathrm{A})}(\sigma_1, \sigma_2) &=& c_{\sigma_2} - c_{\sigma_1} \quad (\forall \sigma_1, \forall \sigma_2)
 \label{eq:difference_game}
\end{eqnarray}
with some function $c_{\sigma}$.
\end{lemma}

\textit{Proof.}
For two-player symmetric game, the definition of potential game is explicitly written as
\begin{eqnarray}
 s_1(\sigma_1, \sigma_2) - s_1(\sigma^\prime_1, \sigma_2) &=& \Phi(\sigma_1, \sigma_2) - \Phi(\sigma^\prime_1, \sigma_2) \quad (\forall \sigma_1, \forall \sigma^\prime_1, \forall \sigma_2) \label{eq:potential_1} \\
 s_2(\sigma_1, \sigma_2) - s_2(\sigma_1, \sigma^\prime_2) &=& \Phi(\sigma_1, \sigma_2) - \Phi(\sigma_1, \sigma^\prime_2) \quad (\forall \sigma_2, \forall \sigma^\prime_2, \forall \sigma_1). \label{eq:potential_2}
\end{eqnarray}
Because the game is symmetric, the condition (\ref{eq:potential_2}) is equivalent to
\begin{eqnarray}
 s_1(\sigma_2, \sigma_1) - s_1(\sigma^\prime_2, \sigma_1) &=& \Phi(\sigma_1, \sigma_2) - \Phi(\sigma_1, \sigma^\prime_2).
\end{eqnarray}
By relabeling the name of variables, it is rewritten as
\begin{eqnarray}
 s_1(\sigma_1, \sigma_2) - s_1(\sigma^\prime_1, \sigma_2) &=& \Phi(\sigma_2, \sigma_1) - \Phi(\sigma_2, \sigma^\prime_1).
\end{eqnarray}
Then we obtain
\begin{eqnarray}
 \Phi(\sigma_1, \sigma_2) - \Phi(\sigma^\prime_1, \sigma_2) &=& \Phi(\sigma_2, \sigma_1) - \Phi(\sigma_2, \sigma^\prime_1),
\end{eqnarray}
or
\begin{eqnarray}
 \Phi^{(\mathrm{A})}(\sigma_1, \sigma_2) &=& \Phi^{(\mathrm{A})}(\sigma^\prime_1, \sigma_2),
\end{eqnarray}
which means that the anti-symmetric part of the potential $\Phi$ does not depend on $\sigma_1$.
By using the same argument, we also obtain
\begin{eqnarray}
 \Phi^{(\mathrm{A})}(\sigma_1, \sigma_2) &=& \Phi^{(\mathrm{A})}(\sigma_1, \sigma^\prime_2).
\end{eqnarray}
Therefore, $\Phi^{(\mathrm{A})}$ must be constant.
However, because $\Phi^{(\mathrm{A})}$ is the anti-symmetric part, $\Phi^{(\mathrm{A})}(\sigma, \sigma) = 0$ for any $\sigma$, and this constant must be zero.
Thus, $\Phi^{(\mathrm{A})}(\sigma_1, \sigma_2) = 0$ for all $(\sigma_1, \sigma_2)$, and we conclude that the potential $\Phi$ is symmetric.

By using this fact, we find that
\begin{eqnarray}
 s_1(\sigma_1, \sigma_2) - s_1(\sigma_2, \sigma_1) &=& s_1(\sigma_1, \sigma_2) - s_2(\sigma_1, \sigma_2) \nonumber \\
 &=& \left[ \Phi(\sigma_1, \sigma_2) - \Phi(1, \sigma_2) + s_1(1, \sigma_2) \right] - \left[ \Phi(\sigma_1, \sigma_2) - \Phi(\sigma_1, 1) + s_2(\sigma_1, 1) \right] \nonumber \\
 &=& \left[ - \Phi(1, \sigma_2) + s_1(1, \sigma_2) \right] - \left[ - \Phi(1, \sigma_1) + s_1(1, \sigma_1) \right] \nonumber \\
 &=& d_{\sigma_2} - d_{\sigma_1},
\end{eqnarray}
where we have defined
\begin{eqnarray}
 d_\sigma &:=& - \Phi(1, \sigma) + s_1(1, \sigma).
\end{eqnarray}
Therefore, we obtain the form (\ref{eq:difference_game}).

Conversely, when the condition (\ref{eq:difference_game}) holds, 
\begin{eqnarray}
 s_1(\sigma_1, \sigma_2) - c_{\sigma_2} &=& s_1(\sigma_2, \sigma_1) - c_{\sigma_1}.
\end{eqnarray}
When we introduce the quantity
\begin{eqnarray}
 \phi(\sigma_1, \sigma_2) &:=& s_1(\sigma_1, \sigma_2) - c_{\sigma_2},
\end{eqnarray}
it satisfies
\begin{eqnarray}
 \phi(\sigma_1, \sigma_2) &=& \phi(\sigma_2, \sigma_1).
\end{eqnarray}
Then we find that
\begin{eqnarray}
 s_1(\sigma_1, \sigma_2) - s_1(\sigma^\prime_1, \sigma_2) &=& \left[ \phi(\sigma_1, \sigma_2) + c_{\sigma_2} \right] - \left[ \phi(\sigma^\prime_1, \sigma_2) + c_{\sigma_2} \right] \nonumber \\
 &=& \phi(\sigma_1, \sigma_2) - \phi(\sigma^\prime_1, \sigma_2)
\end{eqnarray}
and
\begin{eqnarray}
 s_2(\sigma_1, \sigma_2) - s_2(\sigma_1, \sigma^\prime_2) &=& s_1(\sigma_2, \sigma_1) - s_1(\sigma^\prime_2, \sigma_1) \nonumber \\
 &=& \phi(\sigma_2, \sigma_1) - \phi(\sigma^\prime_2, \sigma_1) \nonumber \\
 &=& \phi(\sigma_1, \sigma_2) - \phi(\sigma_1, \sigma^\prime_2).
\end{eqnarray}
Therefore, the quantity $\phi$ can be regarded as a potential.
$\Box$

We now prove our main theorem.
\begin{theorem}
\label{th:ZDTFT}
For two-player symmetric games, TFT is a ZD strategy if and only if the stage game is a potential game.
\end{theorem}

\textit{Proof.}
When player $1$ takes TFT, her Press-Dyson vector is described as
\begin{eqnarray}
 \hat{T}_1\left( \sigma_1 | \sigma^\prime_1, \sigma^\prime_2 \right) &=& \delta_{\sigma_1, \sigma^\prime_2} - \delta_{\sigma_1, \sigma^\prime_1}.
\end{eqnarray}
If TFT is a ZD strategy (\ref{eq:ZDS}), it satisfies
\begin{eqnarray}
 c_{\sigma^\prime_2} - c_{\sigma^\prime_1} &=& \sum_{b=1}^2 \alpha_{b} s_{b} \left( \sigma^\prime_1, \sigma^\prime_2 \right) + \alpha_0
 \label{eq:ZDS_TFT}
\end{eqnarray}
with some non-trivial coefficients $\left\{ \alpha_{b} \right\}$ and $\left\{ c_{\sigma_a} \right\}$.
Because
\begin{eqnarray}
 c_{\sigma^\prime_1} - c_{\sigma^\prime_2} &=& \sum_{b=1}^2 \alpha_{b} s_{b} \left( \sigma^\prime_2, \sigma^\prime_1 \right) + \alpha_0,
\end{eqnarray}
we obtain
\begin{eqnarray}
 -\sum_{b=1}^2 \alpha_{b} s_{b} \left( \sigma^\prime_1, \sigma^\prime_2 \right) - \alpha_0 &=& \sum_{b=1}^2 \alpha_{b} s_{b} \left( \sigma^\prime_2, \sigma^\prime_1 \right) + \alpha_0,
\end{eqnarray}
or
\begin{eqnarray}
 0 &=& \sum_{b=1}^2 \alpha_{b} s^{(\mathrm{S})}_{b} \left( \sigma^\prime_1, \sigma^\prime_2 \right) + \alpha_0,
\end{eqnarray}
for all $\left( \sigma^\prime_1, \sigma^\prime_2 \right)$.
Furthermore, because the game is symmetric, this equation can be rewritten as
\begin{eqnarray}
 0 &=& \left( \alpha_1 + \alpha_2 \right) s^{(\mathrm{S})}_1 \left( \sigma^\prime_1, \sigma^\prime_2 \right) + \alpha_0,
\end{eqnarray}
Then, the coefficients must satisfy either of the following relations:
\begin{eqnarray}
 \left\{
  \begin{array}{ll}
    \alpha_0 = - \left( \alpha_1 + \alpha_2 \right) s^{(\mathrm{S})}_1 \left( 1, 1 \right) & \left( \mathrm{if} \quad s^{(\mathrm{S})}_1 \left( \sigma^\prime_1, \sigma^\prime_2 \right) = s^{(\mathrm{S})}_1 \left( 1, 1 \right) \quad (\forall \sigma^\prime_1, \forall \sigma^\prime_2) \right)\\
   \alpha_1 + \alpha_2 = 0, \quad \alpha_0 = 0 & (\mathrm{otherwise}).
  \end{array}
  \right.
\end{eqnarray}
For the former case, Eq. (\ref{eq:ZDS_TFT}) can be rewritten as
\begin{eqnarray}
 c_{\sigma^\prime_2} - c_{\sigma^\prime_1} &=& \sum_{b=1}^2 \alpha_{b} s_{b} \left( \sigma^\prime_1, \sigma^\prime_2 \right) - \left( \alpha_1 + \alpha_2 \right) s^{(\mathrm{S})}_1 \left( 1, 1 \right) \nonumber \\
 &=& \alpha_1 \left[ s^{(\mathrm{S})}_1 \left( \sigma^\prime_1, \sigma^\prime_2 \right) + s^{(\mathrm{A})}_1 \left( \sigma^\prime_1, \sigma^\prime_2 \right) \right] + \alpha_2 \left[ s^{(\mathrm{S})}_1 \left( \sigma^\prime_2, \sigma^\prime_1 \right) + s^{(\mathrm{A})}_1 \left( \sigma^\prime_2, \sigma^\prime_1 \right) \right] - \left( \alpha_1 + \alpha_2 \right) s^{(\mathrm{S})}_1 \left( 1, 1 \right) \nonumber \\
 &=& \alpha_1 s^{(\mathrm{A})}_1 \left( \sigma^\prime_1, \sigma^\prime_2 \right) + \alpha_2 s^{(\mathrm{A})}_1 \left( \sigma^\prime_2, \sigma^\prime_1 \right) \nonumber \\
 &=& (\alpha_1-\alpha_2) s^{(\mathrm{A})}_1 \left( \sigma^\prime_1, \sigma^\prime_2 \right).
\end{eqnarray}
For the latter case, Eq. (\ref{eq:ZDS_TFT}) can be rewritten as
\begin{eqnarray}
 c_{\sigma^\prime_2} - c_{\sigma^\prime_1} &=& \alpha_1 s_1 \left( \sigma^\prime_1, \sigma^\prime_2 \right) - \alpha_1 s_2 \left( \sigma^\prime_1, \sigma^\prime_2 \right) \nonumber \\
 &=& \alpha_1 s_1 \left( \sigma^\prime_1, \sigma^\prime_2 \right) - \alpha_1 s_1 \left( \sigma^\prime_2, \sigma^\prime_1 \right) \nonumber \\
 &=& 2\alpha_1 s^{(\mathrm{A})}_1 \left( \sigma^\prime_1, \sigma^\prime_2 \right).
\end{eqnarray}
Therefore, $c_{\sigma^\prime_2} - c_{\sigma^\prime_1}$, which is non-zero for some pairs $\left( \sigma^\prime_1, \sigma^\prime_2 \right)$, is proportional to $s^{(\mathrm{A})}_1 \left( \sigma^\prime_1, \sigma^\prime_2 \right)$ for both cases.
From Lemma \ref{lemma:potential}, this is the condition for a game to be a potential game.
Therefore, if TFT is a ZD strategy, then the game must be a potential game.

Conversely, if the game is a potential game, it satisfies
\begin{eqnarray}
 c_{\sigma^\prime_2} - c_{\sigma^\prime_1} &=& s^{(\mathrm{A})}_1 \left( \sigma^\prime_1, \sigma^\prime_2 \right)
\end{eqnarray}
for some function $c_{\sigma}$ (Lemma \ref{lemma:potential}).
This means that
\begin{eqnarray}
 c_{\sigma^\prime_2} - c_{\sigma^\prime_1} &=& \frac{1}{2} s_1 \left( \sigma^\prime_1, \sigma^\prime_2 \right) - \frac{1}{2} s_1 \left( \sigma^\prime_2, \sigma^\prime_1 \right) \nonumber \\
 &=& \frac{1}{2} s_1 \left( \sigma^\prime_1, \sigma^\prime_2 \right) - \frac{1}{2} s_2 \left( \sigma^\prime_1, \sigma^\prime_2 \right)
\end{eqnarray}
with
\begin{eqnarray}
 c_{\sigma^\prime_2} - c_{\sigma^\prime_1} &=& \sum_{\sigma_1} c_{\sigma_1} \hat{T}_1\left( \sigma_1 | \sigma^\prime_1, \sigma^\prime_2 \right).
\end{eqnarray}
Therefore, if the game is a potential game, TFT is a ZD strategy.
$\Box$

The following corollary is a direct consequence of Theorem \ref{th:ZDTFT} and Proposition \ref{prop:ZDS}.
\begin{corollary}
\label{cor:TFT_linear}
For two-player symmetric games, TFT unilaterally enforces
\begin{eqnarray}
 \left\langle s_1 \right\rangle^{*} &=& \left\langle s_2 \right\rangle^{*}
 \label{eq:TFT_linear}
\end{eqnarray}
for potential games.
\end{corollary}

\textit{Proof.}
In the proof of Theorem \ref{th:ZDTFT}, we find that
\begin{eqnarray}
 \sum_{\sigma_1} c_{\sigma_1} \hat{T}_1\left( \sigma_1 | \sigma^\prime_1, \sigma^\prime_2 \right) &=& \frac{1}{2} s_1 \left( \sigma^\prime_1, \sigma^\prime_2 \right) - \frac{1}{2} s_2 \left( \sigma^\prime_1, \sigma^\prime_2 \right)
\end{eqnarray}
for potential games.
By using Proposition \ref{prop:ZDS}, we obtain Eq. (\ref{eq:TFT_linear}).
$\Box$

We remark that the linear relation enforced by TFT in potential games is restricted to Eq. (\ref{eq:TFT_linear}).
Therefore, TFT can only unilaterally equalizes the expected payoffs of two players in potential games.

\subsection{TFT in non-potential games}
\label{subsec:linear}
Theorem \ref{th:ZDTFT} claims that TFT cannot be a ZD strategy in non-potential games.
However, this does not directly lead to the statement that TFT cannot unilaterally enforce any linear relations between expected payoffs in non-potential games, since there is no guarantee that ZD strategies are only strategies which unilaterally enforce linear relations between expected payoffs.
In Ref. \cite{MamIch2019}, the authors proved that memory-one strategies which unilaterally enforce linear relations between expected payoffs are restricted to ZD strategies and unconditional strategies in the prisoner's dilemma game, if both players use memory-one strategies and a stationary distribution of the induced Markov chain exists.
Here we extend their results to our case.

\begin{theorem}
\label{th:linear}
For two-player symmetric games, if the opponent $-a$ also uses a time-independent memory-one strategy $T_{-a}\left( \sigma_{-a} | \bm{\sigma}^{\prime} \right)$ and the induced Markov chain
\begin{eqnarray}
 P_{t+1}\left( \bm{\sigma} \right) &=& \sum_{\bm{\sigma}^{\prime}} \left\{ \prod_{a=1}^2 T_a\left( \sigma_a | \bm{\sigma}^{\prime} \right) \right\} P_t\left( \bm{\sigma}^{\prime} \right)
 \label{eq:MC}
\end{eqnarray}
has a stationary distribution, then the following two conditions are equivalent:
\begin{enumerate}[(a)]
 \item TFT is a ZD strategy.
 \item TFT unilaterally enforces a linear relation between expected payoffs.
\end{enumerate}
\end{theorem}

\textit{Proof.}
((a)$\Rightarrow$(b)): 
If TFT is a ZD strategy, Theorem \ref{th:ZDTFT} claims that the stage game is a potential game.
Then, under the assumptions, Corollary \ref{cor:TFT_linear} claims that TFT unilaterally enforces a linear relation (\ref{eq:TFT_linear}).

((b)$\Rightarrow$(a)): 
We first note that the limit distribution $P^*$ coincides with the stationary distribution of Eq. (\ref{eq:MC}) under the assumptions.
We assume that TFT of player 1 unilaterally enforces a linear relation between expected payoffs
\begin{eqnarray}
 0 &=& \alpha_1 \left\langle s_1 \right\rangle^{*} + \alpha_2 \left\langle s_2 \right\rangle^{*} + \alpha_0.
 \label{eq:linear_def}
\end{eqnarray}
Due to the assumptions, the stationary distribution exists, such that
\begin{eqnarray}
 P^{*}\left( \bm{\sigma} \right) &=& \sum_{\bm{\sigma}^{\prime}} T\left( \bm{\sigma} | \bm{\sigma}^{\prime} \right) P^{*}\left( \bm{\sigma}^{\prime} \right)
\end{eqnarray}
with the transition probability
\begin{eqnarray}
 T\left( \bm{\sigma} | \bm{\sigma}^{\prime} \right) &:=& \prod_{a=1}^2 T_a\left( \sigma_a | \bm{\sigma}^{\prime} \right).
\end{eqnarray}
By introducing a vector $\bm{P}^{*} := \left( P^{*}\left( \bm{\sigma} \right) \right)_{\bm{\sigma}\in A^2}$ and a matrix $T^\prime := T-\mathbb{I}_{M^2}$, where $\mathbb{I}_k$ is a $k\times k$ identity matrix, this condition can be rewritten as
\begin{eqnarray}
 T^\prime \bm{P}^{*} &=& \bm{0}.
\end{eqnarray}
Therefore, for a non-trivial solution $\bm{P}^{*}$ to exist, $\det{T^\prime}=0$ must hold.
On the other hand, because of the relation between a matrix $T^\prime$ and its adjugate matrix $\mathrm{Adj}(T^\prime)$, we obtain \cite{PreDys2012}
\begin{eqnarray}
 T^\prime \mathrm{Adj}(T^\prime) &=& \left( \det{T^\prime} \right) \mathbb{I}_{M^2} = \bm{O}_{M^2},
\end{eqnarray}
where $\bm{O}_k$ is a $k \times k$ zero matrix.
Then, we find that $\bm{P}^{*}$ is proportional to the every column of $\mathrm{Adj}(T^\prime)$.
By choosing the last column of $\mathrm{Adj}(T^\prime)$ as $\bm{P}^{*}$, we obtain
\begin{eqnarray}
 P^{*}\left( \bm{\sigma} \right) &=& C \cdot \left( \mathrm{Adj}(T^\prime) \right)_{\bm{\sigma}, (M, M)} \quad (\forall \bm{\sigma}),
\end{eqnarray}
where $C$ is a constant.
By using this fact, the expected value of a quantity $B$ with respect to the stationary distribution is
\begin{eqnarray}
 \left\langle B \right\rangle^{*} &=& \sum_{\bm{\sigma}} B(\bm{\sigma}) P^{*}\left( \bm{\sigma} \right) \nonumber \\
 &=& C D(\bm{B}),
\end{eqnarray}
where we have defined
\begin{eqnarray}
 D(\bm{B}) &:=& \left|
\begin{array}{c}
 \bm{T}^\prime(1,1)^\mathsf{T}  \\
\bm{T}^\prime(1,2)^\mathsf{T}  \\
 \vdots  \\
 \bm{T}^\prime(M,M-1)^\mathsf{T} \\
 \bm{B}^\mathsf{T}
\end{array}
\right|,
\end{eqnarray}
and vectors $\bm{T}^\prime(\bm{\sigma}):=\left( T^\prime\left( \bm{\sigma} | \bm{\sigma}^{\prime} \right) \right)_{\bm{\sigma}^\prime \in A^2}$ and $\bm{B} := \left( B \left( \bm{\sigma} \right) \right)_{\bm{\sigma}\in A^2}$.
We find that $C=D(\bm{1})^{-1}$, where $\bm{1}$ is a vector of all ones.
Then, a linear relation (\ref{eq:linear_def}) can be rewritten as
\begin{eqnarray}
 0 &=& \frac{D(\alpha_1 \bm{s}_1 + \alpha_2 \bm{s}_2 + \alpha_0 \bm{1})}{D(\bm{1})}.
 \label{eq:linear_det}
\end{eqnarray}
Below we set $\bm{B}=\alpha_1 \bm{s}_1 + \alpha_2 \bm{s}_2 + \alpha_0 \bm{1}$.

The necessary and sufficient condition for Eq. (\ref{eq:linear_det}) to hold is that $M^2$ vectors $\bm{T}^\prime(1,1)$, $\cdots$, $\bm{T}^\prime(M,M-1)$, $\bm{B}$ are linearly dependent, that is
\begin{eqnarray}
 \sum_{\bm{\sigma}\neq (M,M)} c_{\bm{\sigma}} \bm{T}^\prime(\bm{\sigma}) + c_{(M,M)} \bm{B} &=& \bm{0}
 \label{eq:necsuf_dep}
\end{eqnarray}
for some non-trivial $\{ c_{\bm{\sigma}} \}$ (that is, not $c_{\bm{\sigma}}=0$ $(\forall \bm{\sigma})$).
When player 1 uses TFT, Eq. (\ref{eq:necsuf_dep}) is written as
\begin{eqnarray}
 0 &=& \sum_{\bm{\sigma}\neq (M,M)} c_{\bm{\sigma}} \left[ \delta_{\sigma_1, \sigma^\prime_2} T_2\left( \sigma_2 | \bm{\sigma}^{\prime} \right) - \delta_{\sigma_1, \sigma^\prime_1} \delta_{\sigma_2, \sigma^\prime_2} \right] + c_{(M,M)} B(\bm{\sigma}^\prime) \quad (\forall \bm{\sigma}^\prime).
 \label{eq:necsuf_dep_compo}
\end{eqnarray}
These are $M^2$ simultaneous equations of $M^2$ variables $\{c_{\bm{\sigma}}\}$.
Or, explicitly, the vectors $\bm{T}^\prime( \sigma_1, \sigma_2 )$ are
\begin{eqnarray}
 \bm{T}^\prime( \sigma_1, \sigma_2 ) &=& \left(
\begin{array}{c}
 0  \\
 \vdots  \\
 T_2 \left( \sigma_2 | 1, \sigma_1 \right)  \\
 \vdots \\
 0 \\
 0  \\
 \vdots  \\
 T_2 \left( \sigma_2 | 2, \sigma_1 \right)  \\
 \vdots \\
 0 \\
 \vdots \\
 0 \\
 \vdots  \\
 T_2 \left( \sigma_2 | M, \sigma_1 \right)  \\
 \vdots \\
 0 
\end{array}
\right) - \left(
\begin{array}{c}
 0  \\
 \vdots  \\
 0  \\
 1 \quad (\leftarrow (\sigma_1, \sigma_2)) \\
 0 \\
 \vdots \\
 0 
\end{array}
\right) \quad (\forall \bm{\sigma}\neq (M,M)),
\end{eqnarray}
and therefore Eq. (\ref{eq:necsuf_dep}) can be expressed as
\begin{eqnarray}
 \left(
\begin{array}{c}
 \sum_{\sigma_2} c_{(1, \sigma_2)} T_2\left( \sigma_2 | 1, 1 \right)  \\
 \vdots  \\
 \sum_{\sigma_2} c_{(M-1, \sigma_2)} T_2\left( \sigma_2 | 1, M-1 \right)  \\
 \sum_{\sigma_2\neq M} c_{(M, \sigma_2)} T_2\left( \sigma_2 | 1, M \right) \\
 \vdots \\
 \sum_{\sigma_2} c_{(1, \sigma_2)} T_2\left( \sigma_2 | M, 1 \right)  \\
 \vdots  \\
 \sum_{\sigma_2} c_{(M-1, \sigma_2)} T_2\left( \sigma_2 | M, M-1 \right)  \\
 \sum_{\sigma_2\neq M} c_{(M, \sigma_2)} T_2\left( \sigma_2 | M, M \right) 
\end{array}
\right) - \left(
\begin{array}{c}
 c_{(1,1)}  \\
 \vdots  \\
 c_{(1,M-1)}  \\
 c_{(1,M)} \\
 \vdots \\
 c_{(M,1)}  \\
 \vdots  \\
 c_{(M,M-1)}  \\
 0
\end{array}
\right) + c_{(M,M)} \bm{B} &=& \bm{0}.
\label{eq:necsuf_dep_expl}
\end{eqnarray}
We remark that the normalization condition of $T_2$ leads to
\begin{eqnarray}
 T_2\left( M | \sigma^\prime_1, \sigma^\prime_2 \right) &=& 1 - \sum_{\sigma_2=1}^{M-1} T_2\left( \sigma_2 | \sigma^\prime_1, \sigma^\prime_2 \right) \quad (\forall \bm{\sigma}^\prime).
\end{eqnarray}
By using this fact, Eq. (\ref{eq:necsuf_dep_expl}) is rewritten as
\begin{eqnarray}
 \left(
\begin{array}{c}
 \sum_{\sigma_2\neq M} \left( c_{(1, \sigma_2)} - c_{(1, M)} \right) T_2\left( \sigma_2 | 1, 1 \right)  \\
 \vdots  \\
 \sum_{\sigma_2\neq M} \left( c_{(M-1, \sigma_2)} - c_{(M-1, M)} \right) T_2\left( \sigma_2 | 1, M-1 \right)  \\
 \sum_{\sigma_2\neq M} c_{(M, \sigma_2)} T_2\left( \sigma_2 | 1, M \right) \\
 \vdots \\
 \sum_{\sigma_2\neq M} \left( c_{(1, \sigma_2)} - c_{(1, M)} \right) T_2\left( \sigma_2 | M, 1 \right)  \\
 \vdots  \\
 \sum_{\sigma_2\neq M} \left( c_{(M-1, \sigma_2)} - c_{(M-1, M)} \right) T_2\left( \sigma_2 | M, M-1 \right)  \\
 \sum_{\sigma_2\neq M} c_{(M, \sigma_2)} T_2\left( \sigma_2 | M, M \right) 
\end{array}
\right) + \left(
\begin{array}{c}
 c_{(1,M)}  \\
 \vdots  \\
 c_{(M-1,M)}  \\
 0 \\
 \vdots \\
 c_{(1,M)}  \\
 \vdots  \\
 c_{(M-1,M)}  \\
 0 
\end{array}
\right) - \left(
\begin{array}{c}
 c_{(1,1)}  \\
 \vdots  \\
 c_{(1,M-1)}  \\
 c_{(1,M)} \\
 \vdots \\
 c_{(M,1)}  \\
 \vdots  \\
 c_{(M,M-1)}  \\
 0
\end{array}
\right) + c_{(M,M)} \bm{B} &=& \bm{0}. \nonumber \\
&&
\end{eqnarray}
Since we consider the situation that TFT unilaterally enforces a linear relation between expected payoffs, this equation must hold irrespective of the strategy $T_2$ of player 2.
Therefore, the first vector must be zero.
We remark that the coefficient of $T_2\left( \sigma_2 | i,j \right)$ in these equations is common for all $i$.
This leads to
\begin{eqnarray}
 c_{(i, \sigma_2)} &=& c_{i} \quad (1\leq i \leq M-1, 1\leq \sigma_2 \leq M)
 \label{eq:c_i}
\end{eqnarray}
and
\begin{eqnarray}
 c_{(M, \sigma_2)} &=& 0 \quad (1\leq \sigma_2 \leq M-1).
 \label{eq:c_M}
\end{eqnarray}
Below, we write $c_{M}:=0$.
Substituting Eqs. (\ref{eq:c_i}) and (\ref{eq:c_M}) into Eq. (\ref{eq:necsuf_dep_compo}), we finally obtain for $\forall \bm{\sigma}^\prime$
\begin{eqnarray}
 0 &=& \left( \sum_{\bm{\sigma}}^{\sigma_1\neq M} + \sum_{\bm{\sigma}}^{\sigma_1= M, \sigma_2\neq M} \right) c_{\bm{\sigma}} \left[ \delta_{\sigma_1, \sigma^\prime_2} T_2\left( \sigma_2 | \bm{\sigma}^{\prime} \right) - \delta_{\sigma_1, \sigma^\prime_1} \delta_{\sigma_2, \sigma^\prime_2} \right] + c_{(M,M)} B(\bm{\sigma}^\prime) \nonumber \\
&=& \left( \sum_{\bm{\sigma}}^{\sigma_1\neq M} + \sum_{\bm{\sigma}}^{\sigma_1= M, \sigma_2\neq M} \right) c_{\sigma_1} \left[ \delta_{\sigma_1, \sigma^\prime_2} T_2\left( \sigma_2 | \bm{\sigma}^{\prime} \right) - \delta_{\sigma_1, \sigma^\prime_1} \delta_{\sigma_2, \sigma^\prime_2} \right] + c_{(M,M)} B(\bm{\sigma}^\prime) \nonumber \\
 &=& \sum_{\sigma_1\neq M} c_{\sigma_1} \delta_{\sigma_1, \sigma^\prime_2} - \sum_{\sigma_1\neq M} c_{\sigma_1} \delta_{\sigma_1, \sigma^\prime_1} + c_{(M,M)} B(\bm{\sigma}^\prime) \nonumber \\
 &=& \sum_{\sigma_1} c_{\sigma_1} \delta_{\sigma_1, \sigma^\prime_2} - \sum_{\sigma_1} c_{\sigma_1} \delta_{\sigma_1, \sigma^\prime_1} + c_{(M,M)} B(\bm{\sigma}^\prime) \nonumber \\
 &=& c_{\sigma^\prime_2} - c_{\sigma^\prime_1} + c_{(M,M)} B(\bm{\sigma}^\prime) \nonumber \\
 &=& \sum_{\sigma_1} c_{\sigma_1} \hat{T}_1\left( \sigma_1 | \bm{\sigma}^\prime \right) + c_{(M,M)} B(\bm{\sigma}^\prime).
\end{eqnarray}
Therefore, TFT is a ZD strategy.
$\Box$

Theorem \ref{th:linear} states that TFT cannot unilaterally enforce any linear relations between expected payoffs in non-potential games, if the opponent also uses memory-one strategies and a stationary distribution exists.
Extension of this theorem to the case that the opponent uses memory-$n$ strategies is a subject of future work.

\section{Example}
\label{sec:example}
In this section, we provide two examples of potential game where TFT is a ZD strategy.

\subsection{Two-player three-action game}
We first consider the following two-player three-action symmetric zero-sum game:
\begin{eqnarray}
 \bm{s}_1 &=& \left( 0, -2, -1, 2, 0, 1, 1, -1, 0 \right)^\mathsf{T} \\
 \bm{s}_2 &=& \left( 0, 2, 1, -2, 0, -1, -1, 1, 0 \right)^\mathsf{T}.
\end{eqnarray}
We can easily check that this game is a potential game with $c_1=1/2$, $c_2=-3/2$ and $c_3=-1/2$ in Lemma \ref{lemma:potential}, and the potential is $\bm{\Phi}= \left( 0, 2, 1, 2, 4, 3, 1, 3, 2 \right)^\mathsf{T}$ if we assume that $\Phi(1,1)=0$.
We write the strategy of player $a$ by $\bm{T}_a(\sigma) := \left( T_a\left( \sigma | \bm{\sigma} \right) \right)_{\bm{\sigma}\in A^2}$.
TFT of player $1$ is $\bm{T}_1(1)=(1, 0, 0, 1, 0, 0, 1, 0, 0)^\mathsf{T}$, $\bm{T}_1(2)=(0, 1, 0, 0, 1, 0, 0, 1, 0)^\mathsf{T}$, and $\bm{T}_1(3)=(0, 0, 1, 0, 0, 1, 0, 0, 1)^\mathsf{T}$.
By writing the Press-Dyson vectors as $\bm{\hat{T}}_a(\sigma) := \left( \hat{T}_a\left( \sigma | \bm{\sigma} \right) \right)_{\bm{\sigma}\in A^2}$, we obtain $\bm{\hat{T}}_1(1)=(0, -1, -1, 1, 0, 0, 1, 0, 0)^\mathsf{T}$, $\bm{\hat{T}}_1(2)=(0, 1, 0, -1, 0, -1, 0, 1, 0)^\mathsf{T}$, and $\bm{\hat{T}}_1(3)=(0, 0, 1, 0, 0, 1, -1, -1, 0)^\mathsf{T}$.
We can check that the relation
\begin{eqnarray}
 \sum_{\sigma=1}^3 c_\sigma \bm{\hat{T}}_1(\sigma) &=& \frac{1}{2} \left[ \bm{s}_1 - \bm{s}_2 \right]
\end{eqnarray}
indeed holds, which means that TFT is a ZD strategy.

When player $2$ uses the memory-one strategy $\bm{T}_2(1)=(0, 1, 0, 0, 1, 0, 0, 1, 0)^\mathsf{T}$, $\bm{T}_2(2)=(0, 0, 1, 0, 0, 1, 0, 0, 1)^\mathsf{T}$, and $\bm{T}_2(3)=(1, 0, 0, 1, 0, 0, 1, 0, 0)^\mathsf{T}$, and both players choose the actions in the first round by the uniform probability distribution $(1/3, 1/3, 1/3)$, the time evolution is described by the Markov chain (\ref{eq:MC}) with the transition probability
\begin{eqnarray}
 T &=& \left(
\begin{array}{ccccccccc}
 0 & 0 & 0 & 0 & 0 & 0 & 0 & 0 & 0  \\
 0 & 0 & 0 & 0 & 0 & 0 & 0 & 0 & 0  \\
 1 & 0 & 0 & 1 & 0 & 0 & 1 & 0 & 0  \\
 0 & 1 & 0 & 0 & 1 & 0 & 0 & 1 & 0  \\
 0 & 0 & 0 & 0 & 0 & 0 & 0 & 0 & 0  \\
 0 & 0 & 0 & 0 & 0 & 0 & 0 & 0 & 0  \\
 0 & 0 & 0 & 0 & 0 & 0 & 0 & 0 & 0  \\
 0 & 0 & 1 & 0 & 0 & 1 & 0 & 0 & 1  \\
 0 & 0 & 0 & 0 & 0 & 0 & 0 & 0 & 0
\end{array}
\right)
\end{eqnarray}
and the initial condition $(1/9, 1/9, 1/9, 1/9, 1/9, 1/9, 1/9, 1/9, 1/9)^\mathsf{T}$.
We can easily check that for this initial condition, the Markov chain converges to the stationary distribution $\bm{P}^{*} = (0, 0, 1/3, 1/3, 0, 0, 0, 1/3, 0)^\mathsf{T}$ at the second round.
The expected payoffs are
\begin{eqnarray}
 \left\langle s_1 \right\rangle^{*} = \left\langle s_2 \right\rangle^{*} = 0,
\end{eqnarray}
which is consistent with Corollary \ref{cor:TFT_linear}.

When we consider a slightly different game \cite{DOS2012b}
\begin{eqnarray}
 \bm{s}_1 &=& \left( 0, 0, -1, 0, 0, 1, 1, -1, 0 \right)^\mathsf{T} \\
 \bm{s}_2 &=& \left( 0, 0, 1, 0, 0, -1, -1, 1, 0 \right)^\mathsf{T},
\end{eqnarray}
this game is not a potential game.
When the strategies of both players are the same as those above, we obtain
\begin{eqnarray}
 \left\langle s_1 \right\rangle^{*} &=& -\frac{2}{3} \\
 \left\langle s_2 \right\rangle^{*} &=& \frac{2}{3}.
\end{eqnarray}
Therefore, player $1$ cannot unilaterally enforce a linear relation $\left\langle s_1 \right\rangle^{*} = \left\langle s_2 \right\rangle^{*}$.
In addition, because this game is also a zero-sum game, $\bm{s}_2=-\bm{s}_1$ holds.
Furthermore, it should be noted that $\bm{\hat{T}}_1(3) = - \bm{\hat{T}}_1(1) - \bm{\hat{T}}_1(2)$ holds due to the normalization condition of probability.
Therefore, if TFT is a ZD strategy for this game, the relation
\begin{eqnarray}
 \sum_{\sigma=1}^2 d_\sigma \bm{\hat{T}}_1(\sigma) &=& \alpha_1 \bm{s}_1 + \alpha_0 \bm{1}
\end{eqnarray}
must hold with some non-trivial coefficients.
Since the $(1,1)$ component of the left-hand side is zero, $\alpha_0$ must be zero.
Moreover, since the $(1,3)$ component of $\bm{\hat{T}}_1(2)$ and the $(3,2)$ component of $\bm{\hat{T}}_1(1)$ are zero, $d_1= \alpha_1$ and $d_2 = -\alpha_1$ must hold.
Then the $(1,2)$ component of the left-hand side is $-2\alpha_1$ and that of the right-hand side is zero, leading to contradiction.
Therefore, TFT is not a ZD strategy in this game.

\subsection{Cournot duopoly game}
As noted in Section \ref{sec:preliminaries}, the properties of ZD strategies and potential games hold even if the action set is uncountable.
Moreover, the theoretical results in subsection \ref{subsec:potential} can also be easily extended to the case that the action is a continuous variable.
Therefore, we here consider the Cournot duopoly game with unbounded payoffs.
The action space of both players is $A=[0, \infty)$.
The payoff of player $a$ is given by
\begin{eqnarray}
 s_a \left( \bm{\sigma} \right) &=& \left[ A - B\sum_{b=1}^2 \sigma_b \right] \sigma_a - C \sigma_a.
 \label{eq:payoff_cournot}
\end{eqnarray}
The Cournot duopoly game has a potential
\begin{eqnarray}
 \Phi \left( \bm{\sigma} \right) &=& \left[ (A-C) - B\sum_{b=1}^2 \sigma_b \right] \sum_{b=1}^2 \sigma_b + B \sigma_1 \sigma_2.
 \label{eq:potential_cournot}
\end{eqnarray}
For this continuous action space, the definition (\ref{eq:potential}) of a potential leads to
\begin{eqnarray}
 \frac{\partial s_a}{\partial \sigma_a} \left( \bm{\sigma} \right) &=& \frac{\partial \Phi}{\partial \sigma_a} \left( \bm{\sigma} \right) \quad (\forall \bm{\sigma})
\end{eqnarray}
for all $a$.
We can easily check that this relation indeed holds for Eqs. (\ref{eq:payoff_cournot}) and (\ref{eq:potential_cournot}).

If player $a$ takes TFT
\begin{eqnarray}
 T_a\left( \sigma_a | \bm{\sigma}^{\prime} \right) &=& \delta\left( \sigma_a - \sigma^\prime_{-a} \right),
\end{eqnarray}
where $\delta$ represents the Dirac delta function, her Press-Dyson vectors are
\begin{eqnarray}
 \hat{T}_a\left( \sigma_a | \bm{\sigma}^{\prime} \right) &=& \delta\left( \sigma_a - \sigma^\prime_{-a} \right) - \delta\left( \sigma_a - \sigma^\prime_{a} \right).
\end{eqnarray}
When we consider the quantity
\begin{eqnarray}
 c\left( \sigma \right) &=& \left[ (A-C) - B\sigma \right] \sigma,
\end{eqnarray}
we obtain
\begin{eqnarray}
 \int d\sigma c\left( \sigma \right) \hat{T}_a\left( \sigma | \bm{\sigma}^{\prime} \right) &=& c\left( \sigma^\prime_{-a} \right) - c\left( \sigma^\prime_{a} \right) \nonumber \\
 &=& \left[ (A-C) - B\sigma^\prime_{-a} \right] \sigma^\prime_{-a} - \left[ (A-C) - B\sigma^\prime_{a} \right] \sigma^\prime_{a} \nonumber \\
 &=& \left[ (A-C) - B\sum_{b=1}^2 \sigma^\prime_{b} \right] \sigma^\prime_{-a} - \left[ (A-C) - B\sum_{b=1}^2 \sigma^\prime_{b} \right] \sigma^\prime_{a} \nonumber \\
 &=& s_{-a} \left( \bm{\sigma}^\prime \right) - s_{a} \left( \bm{\sigma}^\prime \right).
\end{eqnarray}
Therefore, TFT is a ZD strategy, and unilaterally enforces $\left\langle s_1 \right\rangle^{*} = \left\langle s_2 \right\rangle^{*}$.

\section{Discussion}
\label{sec:discussion}
In Ref. \cite{DOS2014}, the authors proved that TFT is unbeatable if and only if the stage game is a potential game.
In this section, we introduce their results and discuss the relation between our results and their results.

We first introduce the unbeatable property \cite{DOS2014}.
\begin{definition}
\label{def:unbeatable}
The strategy of player $a$ is \emph{unbeatable} if 
\begin{eqnarray}
 \sum_{t=1}^T \left[ s_{-a}\left( \bm{\sigma}^{(t)} \right) - s_a\left( \bm{\sigma}^{(t)} \right) \right] &\leq& \max_{\bm{\sigma}} \left[ s_{-a}\left( \bm{\sigma} \right) - s_a\left( \bm{\sigma} \right) \right] \quad (\forall T\geq 1)
\end{eqnarray}
for any strategies $\left\{ T_{-a}^{(t)} \left( \sigma_{-a}^{(t)} | h_{[1:t-1]} \right) \right\}_{t=1}^\infty$ of player $-a$.
\end{definition}
Duersch et al. proved the following proposition.
\begin{proposition}[\cite{DOS2014}]
\label{prop:unbeatable}
For two-player symmetric games, TFT is unbeatable if and only if the stage game $G$ is a potential game.
\end{proposition}
When combined with our results, the following three conditions are equivalent for infinitely repeated two-player symmetric games: (i) The stage game is a potential game, (ii) TFT is unbeatable, (iii) TFT is a ZD strategy which unilaterally enforces $\mathcal{S}_1=\mathcal{S}_2$.
Particularly, TFT is unbeatable if and only if TFT is a ZD strategy.

A natural question is whether the equivalence of unbeatable property and a ZD strategy also holds for other strategies including imitation strategies in potential games.
Generally, unbeatable property is easily interpreted and observed in other strategies \cite{DOS2012b}, but extension to other games may be difficult, since difference of payoffs of two players is not always important for other games.
On the other hand, the concept of ZD strategy is clearly defined in general games.
Clearly, ZD strategies are not necessarily unbeatable, as we can see for the case of an equalizer strategy \cite{PreDys2012}.
Meanwhile, here we consider the \emph{imitate-if-better} strategy of player $a$ \cite{DOS2012b}
\begin{eqnarray}
 T_a\left( \sigma_a | \bm{\sigma}^{\prime} \right) &=& \delta_{\sigma_a, \sigma^\prime_{-a}} \mathbb{I}\left( s_{-a} \left( \bm{\sigma}^\prime \right) > s_{a} \left( \bm{\sigma}^\prime \right) \right) + \delta_{\sigma_a, \sigma^\prime_{a}} \mathbb{I}\left( s_{-a} \left( \bm{\sigma}^\prime \right) \leq s_{a} \left( \bm{\sigma}^\prime \right) \right) \quad \left( \forall \sigma_a, \forall \bm{\sigma}^{\prime} \right), \nonumber \\
 &&
 \label{eq:IIB}
\end{eqnarray}
where $\mathbb{I}(\cdots)$ is an indicator function that returns $1$ when $\cdots$ holds and $0$ otherwise.
This strategy has been known to be unbeatable in potential games \cite{DOS2014}.
By using Lemma \ref{lemma:potential}, we find that
\begin{eqnarray}
 \sum_{\sigma_a} c_{\sigma_a} \hat{T}_a\left( \sigma_a | \bm{\sigma}^{\prime} \right) &=& \sum_{\sigma_a} c_{\sigma_a} \left\{ \delta_{\sigma_a, \sigma^\prime_{-a}} - \delta_{\sigma_a, \sigma^\prime_{a}} \right\} \mathbb{I}\left( s_{-a} \left( \bm{\sigma}^\prime \right) > s_{a} \left( \bm{\sigma}^\prime \right) \right) \nonumber \\
 &=& \left\{ \frac{1}{2} s_{a} \left( \bm{\sigma}^\prime \right) - \frac{1}{2} s_{-a} \left( \bm{\sigma}^\prime \right) \right\} \mathbb{I}\left( s_{-a} \left( \bm{\sigma}^\prime \right) > s_{a} \left( \bm{\sigma}^\prime \right) \right)
\end{eqnarray}
for potential games, where the function $c_\sigma$ is that in Eq. (\ref{eq:difference_game}).
Then, from Lemma \ref{lemma:Akin}, we obtain the following proposition.
\begin{proposition}
\label{prop:IIB_DZD}
For two-player symmetric games, the imitate-if-better strategy (\ref{eq:IIB}) of player $a$ unilaterally enforces
\begin{eqnarray}
 0 &=& \left\langle \left\{  s_{a} \left( \bm{\sigma}^\prime \right) - s_{-a} \left( \bm{\sigma}^\prime \right) \right\} \mathbb{I}\left( s_{-a} \left( \bm{\sigma}^\prime \right) > s_{a} \left( \bm{\sigma}^\prime \right) \right) \right\rangle^{*}
 \label{eq:IIB_linear}
\end{eqnarray}
for potential games.
\end{proposition}
This is not a ZD strategy, but is contained in the class of extended ZD strategies which unilaterally enforce linear relations between conditional expectations of payoffs \cite{Ued2020}.
In fact, Eq. (\ref{eq:IIB_linear}) implies that the probability that the state $\bm{\sigma}^\prime$ such that $s_{-a}(\bm{\sigma}^\prime)>s_{a}(\bm{\sigma}^\prime)$ is realized is zero.
Therefore, although unbeatable strategies are not necessarily ZD strategies, they may be contained in the class of extended ZD strategies.
Further investigation is needed to this topic.

Finally, we discuss a slight difference between Ref. \cite{DOS2014} and our results.
We first introduce the following concept.
\begin{definition}
\label{def:weak_unbeatable}
The strategy of player $a$ is \emph{weakly unbeatable} if 
\begin{eqnarray}
 \left\langle s_{a} \right\rangle^{*} &\geq& \left\langle s_{-a} \right\rangle^{*}
\end{eqnarray}
for any strategies of player $-a$.
\end{definition}
It should be noted that an unbeatable strategy is weakly unbeatable.
In Ref. \cite{HTS2015}, weakly unbeatable strategies are called \emph{competitive} (or \emph{rival}) strategies.
The following corollary is a direct consequence of Proposition \ref{prop:unbeatable}.
\begin{corollary}
\label{cor:weak_unbeatable}
For two-player symmetric games, TFT is weakly unbeatable for potential games.
In other words, when player $a$ uses TFT, then
\begin{eqnarray}
 \left\langle s_{a} \right\rangle^{*} &\geq& \left\langle s_{-a} \right\rangle^{*}.
 \label{eq:TFT_weak_ineq}
\end{eqnarray}
\end{corollary}
This corollary is weaker than Corollary \ref{cor:TFT_linear}, because Corollary \ref{cor:TFT_linear} claims that the equality $\left\langle s_{a} \right\rangle^{*} = \left\langle s_{-a} \right\rangle^{*}$ must hold in the inequality (\ref{eq:TFT_weak_ineq}).
In other words, although TFT is unbeatable, TFT can also never win.
On the other hand, as we can see from Eq. (\ref{eq:IIB_linear}), the imitate-if-better strategy can win.
For example, in the prisoner's dilemma game, when player $-a$ uses the strategy which always cooperates (All-$C$), the imitate-if-better strategy wins if she starts with defection.

\section{Concluding Remarks}
\label{sec:conclusion}
In this paper, we proved that, for infinitely repeated two-player symmetric games, TFT is a ZD strategy, which unilaterally enforces $\left\langle s_1 \right\rangle^{*} = \left\langle s_2 \right\rangle^{*}$, if and only if the stage game is a potential game.
We also proved that, TFT cannot unilaterally enforce any linear relations between expected payoffs in non-potential games, if the opponent also uses memory-one strategies and a stationary distribution of the induced Markov chain exists.
We explicitly showed that TFT is a ZD strategy in the two potential games, that is, a two-player three-action zero-sum game and the Cournot duopoly game.
Furthermore, we proved that the imitate-if-better strategy can be regarded as an extended ZD strategy in potential games.
When combined with the results of Duersch et al. \cite{DOS2014}, which proved that TFT is unbeatable if and only if the stage game is a potential game, TFT is unbeatable if and only if TFT is a ZD strategy.
This result suggests that there may be some relations between unbeatable strategies and ZD strategies.

In this paper, we consider only two-player symmetric games.
Extension of our result to multi-player symmetric games is non-trivial, because TFT in multi-player symmetric games cannot be defined uniquely, although there are many multi-player symmetric potential games.
We would like to investigate whether imitation strategies in multi-player symmetric games are ZD strategies or not in future.

Another subject of future work is whether TFT is efficient in asymmetric games.
Even though the game is asymmetric, imitation would be useful if the payoff of a player is similar to that of other player.
We would like to find the condition in which TFT is useful in asymmetric games.

\begin{acknowledgment}
This study was supported by JSPS KAKENHI Grant Number JP20K19884.
\end{acknowledgment}


\bibliographystyle{jpsj}
\bibliography{unbeatable}

\end{document}